\documentclass[preprint,prd,nofootinbib,showpacs]{revtex4-1}
\usepackage{graphicx}% Include figure files
\usepackage{epsfig}
\usepackage{bm}% bold math
\def\d{{\rm d}}

\begin{document}

\title{Determining neutrino mass hierarchy from electron disappearance at 
a Low energy neutrino factory}

\author{Rupak Dutta${}^{1,2}$}
\email{rupak@iith.ac.in}   
\author{Nita Sinha${}^{1}$}   
\email{nita@imsc.res.in}
\author{Sushant K. Raut${}^{3,4}$}
\email{sushant@prl.res.in}
\affiliation{
${}^1$The Institute of Mathematical Sciences, 
Chennai 600113, India
\\
${}^2$ Indian Institute of Technology Hyderabad, Hyderabad 502205, India
\\
${}^3$ Indian Institute of Technology Bombay, Mumbai 400076, India
\\
${}^4$ Physical Research Laboratory, Ahmedabad 380009, India
}
\preprint{IMSc/2012/10/16}

\begin{abstract}
  Recent measurements of large $\theta_{13}$ by Daya Bay and RENO
  reactor experiments have opened up the possibility of determining
  the neutrino mass hierarchy, i.e. the sign of the mass squared splitting
  $\Delta m_{31}^2$, the CP-violating phase $\delta_{CP}$ and the
  octant of $\theta_{23}$. In the light of this result, we study the
  performance of a low energy neutrino factory (LENF) for
  determination of the mass hierarchy. In particular, we explore the
  potential of the $\nu_e$ and $\bar{\nu}_e$ disappearance channels at
  LENF to determine the neutrino mass hierarchy, that is free from the
  uncertainties arising from the unknown $\delta_{CP}$ phase and the
  $\theta_{23}$ octant. We find that using these electron
  neutrino (antineutrino) disappearance channels with a standard LENF,
  it is possible to exclude the wrong hierarchy at $5 \sigma$ with only 2 years of running.
%We further emphasize that this may be feasible even with very low muon beam energies and hence perhaps feasible with the proposed VLENF of a NUSTORM, coupled with an additional far detector.
\end{abstract}

 % 	Neutrino interactions

\maketitle
\section{Introduction}
\label{intro}
The discovery of flavour mixing of atmospheric and solar neutrinos in
the golden years of neutrino oscillations (1998-2004) has led to
extensive theoretical and experimental effort in neutrino physics,
worldwide. Precision measurements of flavour mixing parameters in the
lepton sector opens up yet another window in our quest for physics
beyond the Standard Model.

Neutrino oscillation phenomena are described in terms of six
independent parameters namely, three mixing
angles~($\theta_{23},\,\theta_{13},\, \theta_{12}$), two mass squared
differences~($\Delta m_{31}^2,\,\Delta m_{21}^2$) and a CP-violating
phase~($\delta_{CP}$). In the last decade, data from solar and reactor
neutrino experiments have resulted in information on the sign and
magnitude of $\Delta m_{21}^2$ and a precise value of
$\theta_{12}$~\cite{sno,kamland}. The atmospheric parameters $|\Delta
m_{31}^2|$ and $\theta_{23}$ have been measured and their precision
will be increased by T2K~\cite{t2k} and NO$\nu$A~\cite{nova}.  Recent
measurements of $\theta_{13}$ from T2K~\cite{t2kt13},
MINOS~\cite{minos}, and Double Chooz~\cite{DC} indicated a non zero
value of $\theta_{13}$. This year, a moderately large value of
$\theta_{13}$ has been established by the reactor experiments. Daya
Bay~\cite{daya} claimed that $\theta_{13}$ is non vanishing with a
significance of 5.2 standard deviations while RENO~\cite{reno}
measured a larger central value but with a slightly lower significance
of 4.9$\sigma$ coming from their higher systematic error.  In fact, the
global fits~\cite{Valle, Schwetz} of neutrino oscillation parameters
including these recent measurements exclude $\sin^2\theta_{13} = 0$ at
10.2$\sigma$. Large $\theta_{13}$ enables a wide range of
possibilities for determination of the neutrino mass hierarchy, the
value of $\delta_{CP}$ and the octant of $\theta_{23}$ -- the
remaining unknown neutrino oscillation parameters. In fact, a non zero
value of $\theta_{13}$ is a prerequisite to probe these unknowns.

Neutrino mass hierarchy or the neutrino mass ordering has profound
theoretical implications. Till now we have been unable to resolve the
neutrino mass hierarchy, i.e, whether the hierarchy is normal~(NH,
$\Delta m_{31}^2 > 0$) or inverted~(IH, $\Delta m_{31}^2 < 0$). For
neutrinos passing through long baselines, the effect on oscillation
parameters in presence of earth matter~\cite{msw1,msw2,msw3} is
dependent on the sign of $\Delta m_{31}^2$, hence such long baseline
experiments can resolve the question of mass hierarchy. Moreover, the
effect of the matter potential (enhancement or suppression) for each
hierarchy is different for neutrinos versus antineutrinos, requiring
charge identification of the final lepton produced in the interaction
of the neutrinos (antineutrinos) with the detector material. These
earth matter effects are enhanced if $\theta_{13}$ is sufficiently
large, improving the chances of discovery of mass hierarchy.

With the favourable large $\theta_{13}$, future atmospheric neutrino
experiments will play a crucial role in determination of mass
hierarchy through disappearance channels (muon and
electron). Preliminary studies of the India-based Neutrino Observatory
(INO)~\cite{ino} indicate that, INO with its charge identification
capability of muons, will be able to make a measurement at $2.7\sigma$
statistical significance, with $500\,{\rm kton}$-year
exposure~\cite{choubey}. Similar sensitivity will be possible at
Hyperkamiokande~\cite{hk} (with electron events) and with PINGU at
Icecube~\cite{pingu}, while better sensitivity will be achievable with
a rather futuristic magnetized liquid Argon detector~\cite{Barger}
with visibility of both muon and electron events.  Accelerator long
baseline experiments like T2K and NO$\nu$A will have good sensitivity
but only for some fraction of the $\delta_{CP}$ range. In fact,
recently in Ref.~\cite{Prakash}, a combined analysis of T2K and
NO$\nu$A has been performed and it is found that mass hierarchy can
only be determined for $-150^{\circ} \le \delta_{CP} \le -30^{\circ}$
if the true hierarchy is NH and $30^{\circ} \le \delta_{CP} \le
150^{\circ}$ if it is IH, indicating the difficulty in achieving this
without knowlege of the true $\delta_{CP}$, unless statistics is
sufficiently high~\cite{sanjibglade}.  However, a detailed study of
neutrino mass hierarchy performed in Ref.~\cite{Blennow} has
demonstrated that the use of atmospheric data from the proposed INO in
conjunction with accelerator beam experiments T2K and NO$\nu$A results
in a 3$\sigma$ determination of hierarchy with a high resolution
$100\,{\rm kton}$ detector for $\sin^22\theta_{13} = 0.09$.  Also, in
Ref.~\cite{Agarwalla}, the authors have shown that mass hierarchy can
be determined for any value of $\delta_{CP}$ from the near resonant
matter effect in the $\nu_{\mu} \to \nu_e$ oscillation channel at
5$\sigma$ with a superbeam with an average neutrino energy of $5\,{\rm
  GeV}$ at Super-Kamiokande that is at a distance of $8770\,{\rm km}$
from the proposed superbeam facility at CERN.

In this paper we wish to focus on the proposed low energy neutrino
factory (LENF) for the mass hierarchy determination. In a neutrino
factory both $\nu_{\mu}~(\bar{\nu}_{\mu})$ and $\nu_e~(\bar{\nu}_e)$
beams are produced from the decay of muon~($\mu^{\pm}$) in the long
straight sections of a storage ring. A $\mu^+$ decay produces
$\bar{\nu}_{\mu}$ and $\nu_e$ whereas a $\mu^-$ decays to $\nu_{\mu}$
and $\bar{\nu}_e$. Hence a detailed study of various oscillation
channels such as $\nu_{\mu} \to \nu_e$, $\nu_{\mu} \to \nu_{\mu}$,
$\nu_e \to \nu_{\mu}$, $\nu_e \to \nu_e$ and all the corresponding
antineutrino oscillations are possible using a neutrino factory. In
addition there are the tau appearance channels, however, these can be
utilized only in the presence of a far detector with tau lepton
identification capability. Apart from the availability of large number
of channels, this facility is preferred due to the feasibility of high
beam intensity and accurate knowledge of the neutrino fluxes.

The so called `golden' channel, i.e, the $\nu_e \to \nu_{\mu}$
oscillation, contains information on all the oscillation parameters
and hence in principle can be used to determine all the unknown
parameters. However, degenerate solutions in the parameter space make
this task difficult. Till recently, since the unknowns were:
$\theta_{13}$, CP violating phase $\delta_{CP}$, the octant of
$\theta_{23}$ (if nonmaximal) and the mass hierarchy, this resulted in
an eight fold parameter degeneracy~\cite{Minakata,degenbarger}, making
it hard to extract unique values of the parameters. The solution to
this problem was first provided in Ref.~\cite{degenbarger,magic}.
They found that at the `magic' baseline of around $7500~{\rm km}$, the
$\nu_e \to \nu_{\mu}$ oscillation probability does not depend on the
CP phase $\delta_{CP}$ irrespective of the beam energy. Indeed it
would be possible to determine the mass hierarchy at the magic
baseline and the high energy neutrino factory (HENF) would be the
suitable facility for this long baseline study. The physics potential
of a HENF has been studied by various
authors~\cite{various_henf_papers} and a detail analysis of
optimization of the neutrino factory baseline by the IDS-NF group
suggests two baselines: one at the magic baseline $L = 7000 -
8000\,{\rm km}$ and the other at $L = 2500 - 5000\,{\rm km}$ with a
muon beam energy of $25\,{\rm GeV}$~\cite{ids-nf}. Recent studies
however suggested that a LENF with a baseline of $1300\,{\rm
  km}$~(FNAL to DUSEL) with a muon beam energy of $4.5\,{\rm GeV}$
could achieve the desired precision if $\theta_{13}$ is large
enough~\cite{Geer-Mena}. A re-optimization of the neutrino factory
resulted in the proposal for a staged approach~\cite{Agarwalla-Huber},
which may have the option of much lower ~($4 -10~{\rm GeV}$) muon beam
energies and shorter baselines~($1000-2500\,{\rm km}$). The LENF has
already been shown to have excellent sensitivity to the mass hierarchy
from the golden and the platinum, $\nu_{\mu} \to \nu_e$
channels~\cite{Martinez}. In Ref.~\cite{Dighe}, it has been shown that
only if $\sin^2\theta_{13} > 10^{-2}$, mass hierarchy can be
determined at 5$\sigma$ independently of $\delta_{CP}$ at muon beam energies more than $5\,{\rm GeV}$ and baselines longer than $1000\,{\rm km}$, using the golden channel.

In this study, we show that, with the knowledge of $\theta_{13}$ from
the recent reactor experiments, determination of mass hierarchy can be
made completely free of degeneracies by exploiting the electron
disappearance channel. For electron disappearance ($\nu_e/\bar{\nu}_e
\to \nu_e/\bar{\nu}_e$), the oscillation probability $P_{ee}$ for
neutrinos (antineutrinos) of energy $E_\nu$ traveling over a baseline of
length $L$ in the presence of matter potential $V =\pm\sqrt{2}G_F n_e
$, can be expressed as a perturbative expansion in the parameters
$\alpha = \Delta m_{21}^2/\Delta m_{31}^2$ and $\sin\theta_{13}$
as~\cite{Akhmedov}:
\begin{equation}
  \label{eq:pee}
P_{ee} = 1 - \alpha^2\,\sin^2\,2\theta_{12}\frac{\sin^2A\Delta}{A^2} - 
4\,\sin^2\theta_{13}\,\frac{\sin^2(A-1)\Delta}{(A-1)^2} \,,
\end{equation}
where, $\Delta = (\Delta m_{31}^2\,L/4\,E_\nu)$, $A = (2\,E_\nu\,V/\Delta
m_{31}^2)$, $G_F$ is the Fermi constant and $n_e$ is the electron
density. The sign of the potential $V$ is positive for neutrinos and
negative for antineutrinos.  From Eq.~\ref{eq:pee}, one can see that
since this oscillation channel is independent of $\delta_{CP}$ and
$\theta_{23}$, it is free from the degeneracies arising from the
unkown octant of $\theta_{23}$ and from the lack of knowledge of the
CP violating phase. Hence, at least in principle, it is possible to
use this channel to determine the hierarchy independently of both
these parameter values.

Previous studies of the neutrino factories performance had ignored
this channel. The reason being that one assumed that $\theta_{13}$ is
perhaps very small. Mass hierarchy sensitivity of a setup was then
judged by the smallest value of $\sin^2 2\theta_{13}$ above which the
wrong hierarchy could be excluded. In $P_{ee}$, since the mass
hierarchy sensitive term is quadratic in $\sin\theta_{13}$, therefore
one did not expect much sensitivity for small reactor angles. On the
other hand in the golden channel while there is one atmospheric (mass
hierarchy sensitive) term appearing with $\sin^22\theta_{13}$
dependence and could dominate only for large $1-3$ mixing angle, but
in addition, there is also an interference term which is linear in
$\sin2\theta_{13}$ which allows mass hierarchy sensitivity for
intermediate values also. For very small $\theta_{13}$, sensitivity
was expected only with very high intensity neutrino beams and long
baselines to enhance the matter effect and hence HENFs of $50\,{\rm
  GeV}$ beam energies and rather long baselines were
considered. Feasibility of mass hierarchy determination with these
facilities was shown even for as small values as $\sin^2 2\theta_{13}
\leq 10^{-4}$. In case of muon disappearance, there are many terms
which are sensitive to mass hierarchy, with coefficients independent
of $\theta_{13}$ but quadratic in solar mass squared difference, ones
that depend on $\sin^2\theta_{13}$ as well as those which are linear
in $\sin\theta_{13}$ and the solar mass squared difference. With many
terms of differing signs, for a small value of the reactor angle,
sensitivity was possible again only by going to long baselines and
high energy intense beams, while with large values it is feasible even
with atmospheric neutrinos as described earlier. Apart from the
dependence on the $1-3$ mixing angle, the sensitivity to systematics
for the electron disappearance channel also had a role to play. Any
experiment that measures events coming from $P_{ee}$, a disappearance
channel, has to detect a small deficit in the expected number of
neutrino events. The extent of the deficit depends of course on the
value of $\theta_{13}$. In the light of recent measurements indicating
a large value of $\theta_{13}$, we claim that it is possible to
extract information about the hierarchy from this channel.

Focussing in this paper on the electron disappearance channel at LENF,
we demonstrate that for a large value of $\theta_{13}$~(Daya Bay
range) it is possible to exclude the wrong hierachy at a $5\sigma$ for
all values of $\delta_{CP}$ and any octant of $\theta_{23}$,
irrespective of the choice of the true hierarchy.  Hence unlike all
other oscillation channels, this channel can be used for a clean
determination of the mass hierarchy not just at some magic baselines,
but rather for all baselines greater than about $1200 \,{\rm km}$
corresponding to muon beam energies about $3\,{\rm GeV}$ or
larger. Moreover, this channel can give us an independent confirmation
of hierarchy measurement from the golden channel.  The possibility of
using electron disappearance channel to study mass hierarchy has
been studied in Ref.~\cite{Agarwalla1,peltoniemi,campagne,Donini} in
the context of a $\beta$-beam as a $\nu_e$ source, however, this
requires rather high boost power.

After the recent measurement of a large $\theta_{13}$ value, the
possibilty of determining mass hierarchy with electron disappearance
with reactor neutrinos has also been investigated again by several
authors. Authors of Ref.~\cite{Qian} have concluded that such a
measurement will be difficult due to the finite detector energy
resolution, while Ref.~\cite{Ciuffoli} have described the challenges
and possible solutions. In Ref.~\cite{pomita} it is pointed out that
with rather large exposures, $3\sigma$ mass hierarchy discrimination
seems feasible.

Our paper is organized as follows. In section~\ref{Det}, we start with
a brief description of our experimental setup and then provide a
detail description of the numerical simulations. The results are
reported in section~\ref{res} with a conclusion in section~\ref{con}.

\section{Detector setup and simulations}
\label{Det}
With the large value of $\theta_{13}$ confirmed independently by Daya
Bay and RENO reactor experiments, the LENF is a good facility to
determine the mass hierarchy. A LENF with a baseline of $1300\,{\rm
  km}$ (FNAL to DUSEL) and a muon beam energy of $4.5\,{\rm GeV}$ was
first proposed in Ref.~\cite{Geer-Mena}. It was shown that for sufficiently
high statistics and detection efficiency, an optimized LENF can be an
excellent set up for precision measurements of oscillation parameters
for a large value of $\theta_{13}$. In the context of LENF, two types
of detector technologies have been discussed in the literature: a
$20\,{\rm kton}$ magnetized totally active scintillator detector~(TASD)
and a $100\,{\rm kton}$ liquid argon detector~(LAr) with charge
identification capabilities of both electrons and muons.
 
In our study, we consider an LENF setup of Ref.~\cite{LENF} with a
magnetized $20\,{\rm kton}$ TASD with an energy resolution of $10\%$
for all channels and a lower energy threshold of $0.5\,{\rm GeV}$. The
detection efficiency of the TASD is $37\%$ below and $47\%$ above
$1\,{\rm GeV}$ for electron events with a background at the $10^{-2}$
level. We use a LENF with $1.4\times10^{21}$ useful muons per year
per polarity and a running time of $2$ years. In our initial analysis
we use a baseline of $1300\,{\rm km}$ and muon beam energy $E_{\mu} =
4.5\,{\rm GeV}$.  For more details on low energy neutrino factory we
refer to Ref.~\cite{LENF}.

Our goal is to investigate the performance of the above LENF setup in
determining the neutrino mass hierarchy using electron disappearance
channel, as this channel has the advantage of being independent of
$\delta_{CP}$ and $\theta_{23}$. Earlier, with $\theta_{13}$ unknown
(and hence possibly very tiny), most studies used to look for the
hierarchy reach in $\sin^2 2\theta_{13}$, defining it as the limiting
value of $\sin^2 2\theta_{13}$ above which the wrong hierarchy could
be excluded at a chosen confidence level. Now that $\sin^2
2\theta_{13}$ has been measured to be fairly large with good
precision, there is no point in trying to find the minimum $\sin^2
2\theta_{13}$ for which the wrong heirarchy can be
distinguished. Instead, for most of our analysis, we use the central
value of $\sin^2 2\theta_{13}$ from RENO, as well as the value
corresponding to 2$\sigma$ lower limit of $\sin^2 \theta_{13}$ coming
from the global fit~\cite{Valle} and see if the electron disappearance
channel alone can achieve the desired precision.

For our numerical simulations, we use the following true values of the
oscillation parameters: for the leading atmospheric parameters, we use
$\theta_{23} = 45^{\circ}$ and $\Delta m_{\rm eff}^2 = 2.4\times
10^{-3}\,{\rm eV^2}$, for the solar parameters, we use
$\sin^2\theta_{12}=0.304$ and $\Delta m_{21}^2 = 7.65\times
10^{-5}\,{\rm eV^2}$.  For the third mixing angle $\theta_{13}$, we
use the central value of $\sin^2 2\theta_{13}=0.113$ from RENO.  The
unknown CP violating phase $\delta_{CP}$ is varied over its full range
$-180^{\circ}$ to $180^{\circ}$.  Here $\Delta m^2_{\rm eff}$, an
effective mass-squared difference measured in $\nu_{\mu}$ survival
probability, is related to $\Delta m_{31}^2$ via~\cite{parkedefn}
\begin{eqnarray}
\Delta m^2_{\rm eff} = \Delta m_{31}^2 - (\cos^2\theta_{12} - \cos\delta\sin\theta_{13}\sin2\theta_{12}
\tan\theta_{23})\Delta m_{21}^2\,.
\end{eqnarray}
%The slight $\delta_{CP}$ dependence in the survival
% probability $P_{ee}$ will come from this effective
%$\Delta m_{\rm eff}^2$. We will be using three different 
%definitions of $\Delta m_{31}^2$, namely
%\begin{eqnarray}
%&&\Delta m_{31}^2 = \Delta m_{\rm eff}^2 \nonumber \\
%&&\Delta m_{31}^2 = \Delta m^2_{\rm eff} + (\cos^2\theta_{12} %- \cos\delta\sin\theta_{13}\sin2\theta_{12}
%\cot\theta_{23})\Delta m_{21}^2 \nonumber \\
%&&\Delta m_{31}^2 = \Delta m_{\rm eff}^2 + 
%\frac{\Delta m_{21}^2}{2}
%\end{eqnarray}

To evaluate the event rates for production of electrons in the
detector, we use the differential neutrino factory flux, $\Phi_i
\equiv \d N_i/\d E_\nu$, where $i = e, \mu$\footnote{while we use the
  e-disappearance channels alone for the signal events, however,
  background events appear from the platinum channel also, hence the
  initial muon flux also needs to be considered.} corresponds to
flavour of neutrinos, double differential cross-section $\sigma_e
\equiv \d^2\sigma_e\d E_e/ \d\cos\theta_e$ for CC interactions
(quasi-elastic, resonance and deep inelastic processes) producing an
electron (positron) in the detector and the kinematic constraints that
have been described in detail in Ref.~\cite{Sinha, Dutta}.  The event
rates for production of electrons (positrons) in the detector are
defined as
\begin{eqnarray} \nonumber
{\cal {N}}_{ie} & = & \kappa  \int \Phi_i P_{ie} \sigma_e (\nu_e
\to e) \epsilon_e~,  
\end{eqnarray}
where $\kappa$ accounts for the exposure (size of detector and years of
running), $\epsilon_e$ are the detection efficiencies and the oscillation
probability $P_{ie}$ is a function of the energy $E_\nu$ of the
neutrino and the length of the baseline traveled by the neutrino,
before reaching the detector.
The integration is over all the relevant variables, including
resolution function, corresponding to bins in the observed lepton
energy, $E_e^{\rm obs}$. 
  
For our statistical analysis, we use a gaussian $\chi^2$ with the
events coming from both $\nu_e$ and $\bar{\nu}_e$ disappearance
channels. We generate our experimental data for a fixed or true
hierarchy keeping all other parameters fixed. The theoretical data is
then generated using the other, wrong hierarchy. The resulting
$\chi^2$ thus determines the confidence level at which the wrong
hierarchy can be excluded. However, for realistic analysis one needs
to marginalize over all the parameters in order to include the
uncertainties coming from them. Thus the minimum $\chi^2$ is
determined after taking all the variations into account.  For our
analysis, priors are used for the measured oscillation parameters.  An
error of $10\%$ is taken for the atmospheric parameters $\sin^2
2\theta_{23}$ and $|\Delta m_{\rm eff}^2|$ and $4\%$ on the solar
parameters $\sin^2\theta_{12}$ and $\Delta m_{21}^2$. The error on the
value of $\sin^2\,2\theta_{13}$ is taken to be $0.01$.  Note that
these priors are very conservative. Daya Bay is expected to reduce the
error on $\sin^2 2\theta_{13}$ to 0.005 by 2016. More data from
Borexino~\cite{borexino}, T2K~\cite{t2k} and NO$\nu$A~\cite{nova} will
make more precise measurements of solar and atmospheric parameters,
before the LENF can be constructed.  For the systematics, we have used
normalization error of $2\%$ for the signal and $20\%$ for the
background.  We also include a small tilt of $0.01\%$ for the signal
as well as for the background.  We have included only the charge
mis-identification background, at a level of $10^{-2}$ for the
$\nu_e~(\bar{\nu}_e)$ disappearance channels.

\section{Results and discussion} 
\label{res} 
\begin{figure}[htb]
\begin{center}
\includegraphics[width=10cm,height=7cm]{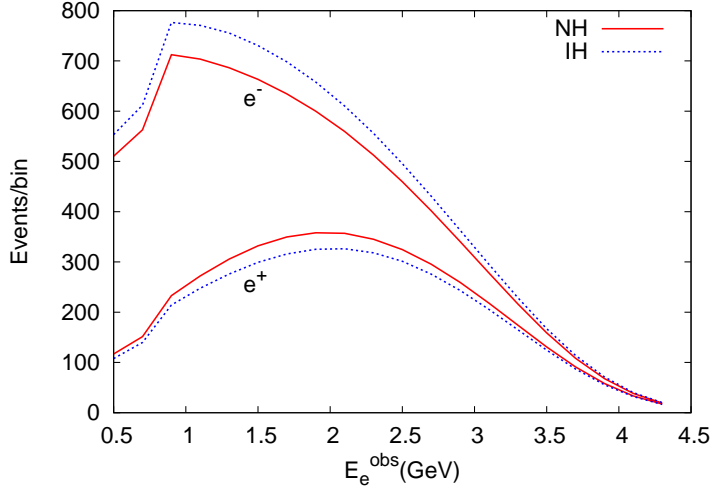}
\end{center}
\caption{\footnotesize{Electron/positron event rates per bin
    ($0.2\,{\rm GeV}$) for $\sin^2 2\theta_{13} = 0.113$, as a
    function of the observed lepton energy at a LENF corresponding to
    NH (red/solid curve) and IH (blue/dashed curve). The events are
    for a $2$ years exposure of a $20\,{\rm kton}$ detector at a
    baseline of $L = 1300\,{\rm km}$, for neutrinos from $4.5\,{\rm
      GeV}$ muon beams with $1.4\times 10^{21}$ useful muon decays per
    year per polarity. The other detector specifications are as
    mentioned in section~\ref{Det}.}}
%threshold, resolution, efficiency and background level and follow the  in Ref.~\cite{LENF}.}} 
\label{events}
\end{figure}

We now proceed to analyse the hierarchy determination ability of
electron disappearance channel using a low energy neutrino factory.
First we fix all the oscillation parameters at their central values
reported in section~\ref{Det} and set $\sin^2 2\theta_{13} = 0.113$
and $\delta_{CP} = 90^{\circ}$. Using the standard LENF setup of muon
beam energy of $4.5\,{\rm GeV}$ and a baseline of $1300\,{\rm km}$,
described in the last section and considering running time of $2$
years, we evaluate the number of electron (positron) events that would
be observed in the detector.  The number of these events resulting
from $\nu_e(\bar{\nu}_e)$ interactions as a function of the observed
lepton energy, corresponding to each of the hierarchies are displayed
in Fig.~\ref{events}. In our entire study we continue to use the same
TASD detector mass, energy resolution, electron detection efficiency,
background level and number of useful muons per polarity, mentioned in
section~\ref{Det} and are as specified in Ref.~\cite{LENF}, even when
the muon beam energy and baseline length are varied.

\begin{figure}[htb]
\begin{center}
\includegraphics[width=0.45\textwidth]{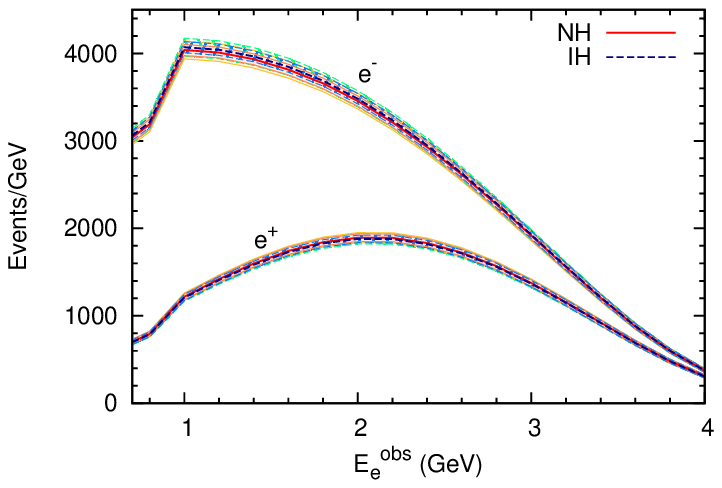}
\includegraphics[width=0.45\textwidth]{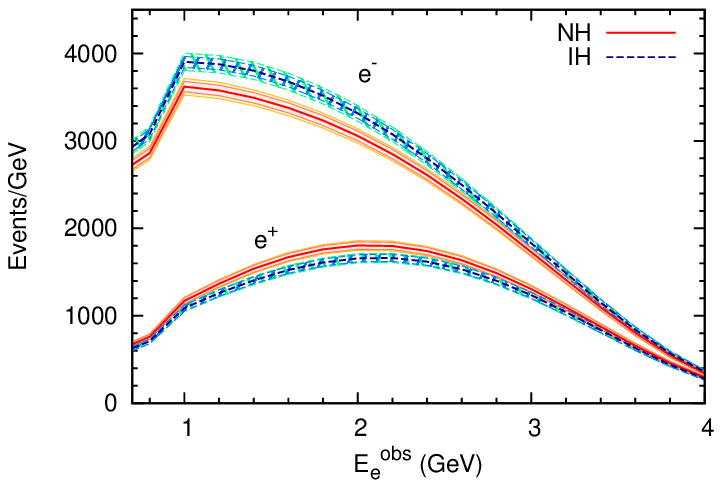}
\end{center}
\caption{\footnotesize{Electron/positron event rates per bin ($1.0\,{\rm GeV}$) for $\sin^2(2\theta_{13}) = 0.01$ 
(Left) and $0.1$ (Right) as a function of the observed lepton energy
at a LENF. The bands around each of the curves denote the combination of statistical and systematic errors added in quadrature. All other specifications are similar to that in Fig.~\ref{events}.}}
\label{events_syst}
\end{figure}

The role of systematics and relative value of $\sin^2 2\theta_{13}$ in
the electron disappearance channel had been discussed in
section~\ref{intro}.  We now substantiate our claim, that for the
current value of $\sin^2 2\theta_{13}$, one can distinguish between
the two hierarchies in spite of systematic uncertainties.  In
Fig.~\ref{events_syst}, we have plotted the $e^-$ and $e^+$ events for
NH and IH for $\sin^2 2\theta_{13} = 0.10$ ($\approx$ current value)
and for $\sin^2 2\theta_{13} =0.01$ (a much smaller value). Bands
around the event rates denote the combination of errors, statistical
and systematic added in quadrature. One can see that for $\sin^2
2\theta_{13} =0.01$, the errors are larger than the difference between
the event rates for the two hierarchies. Even if the exposure and
hence statistics is increased by say a factor of 10, reducing the
relative statistical error, systematic effects will still result in an
overlap between NH and IH. Therefore hierarchy discrimination is not
possible. However, for $\sin^2 2\theta_{13} =0.10$, we see that there
is enough separation between the events corresponding to the two
hierarchies. Hence, with large enough $1-3$ mixing angle, it is
possible to go beyond the regime where electron disappearance had been
a systematics-riddled channel.

\begin{figure}[htb]
\begin{center}
\includegraphics[width=10cm]{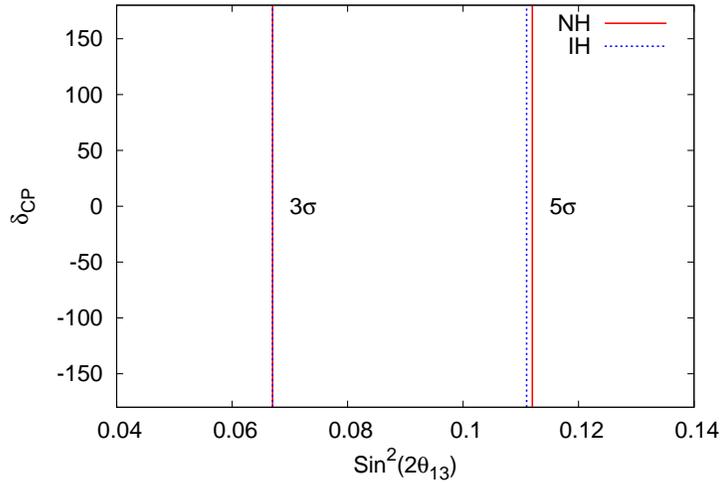}
\end{center}
\caption{\footnotesize{$3\sigma$ and $5\sigma$ Hierarchy exclusion plot 
at a LENF with $4.5\,{\rm GeV}$ $\mu^\pm$ beams (both polarities) over a
2 years exposure with a $20$ kton detector located at a baseline of
$L=1300~\,{\rm km}$.}}
\label{hier}
\end{figure}

For this LENF setup, we determine the exclusion region in the true
value of $\theta_{13}$ and true value of $\delta_{CP}$ plane, for
which the wrong hierarchy can be excluded at $3\sigma$ and
$5\sigma$. Fig.~\ref{hier} exhibits this region and is to be
interpreted as being on the right of the contours. These results have
been obtained after marginalization over all the oscillation
parameters and shown for both normal and inverted hierarchies. We find
that hierarchy can be determined at all values of $\delta_{CP}$,
leading to the contours being vertical lines. It is clear from
Fig.~\ref{hier} that if the true hierarchy is NH, then above $\sin^2
2\theta_{13} = 0.112$ the wrong hierarchy, i.e, IH can be excluded at
$5\sigma$. If IH is the true hierarchy, then NH can be eliminated
at $5\sigma$  for $\sin^2 2\theta_{13} = 0.111$ and beyond. Also,
above $\sin^2 2\theta_{13} = 0.067$, the wrong hierarchy can be
excluded at $3\sigma$, irrespective of the true hierarchy.

\begin{figure}[htb]
\begin{center}
\includegraphics[width=0.45\textwidth]{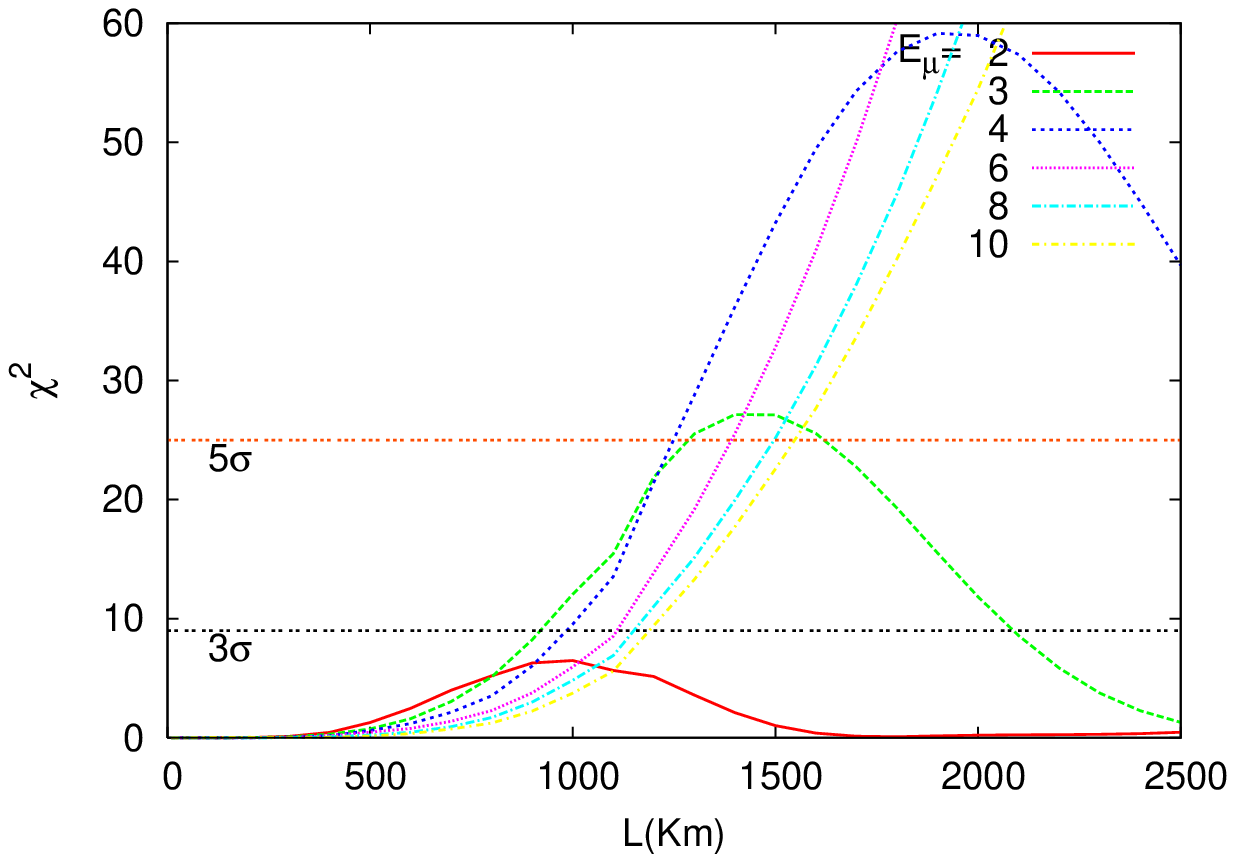}
\includegraphics[width=0.45\textwidth]{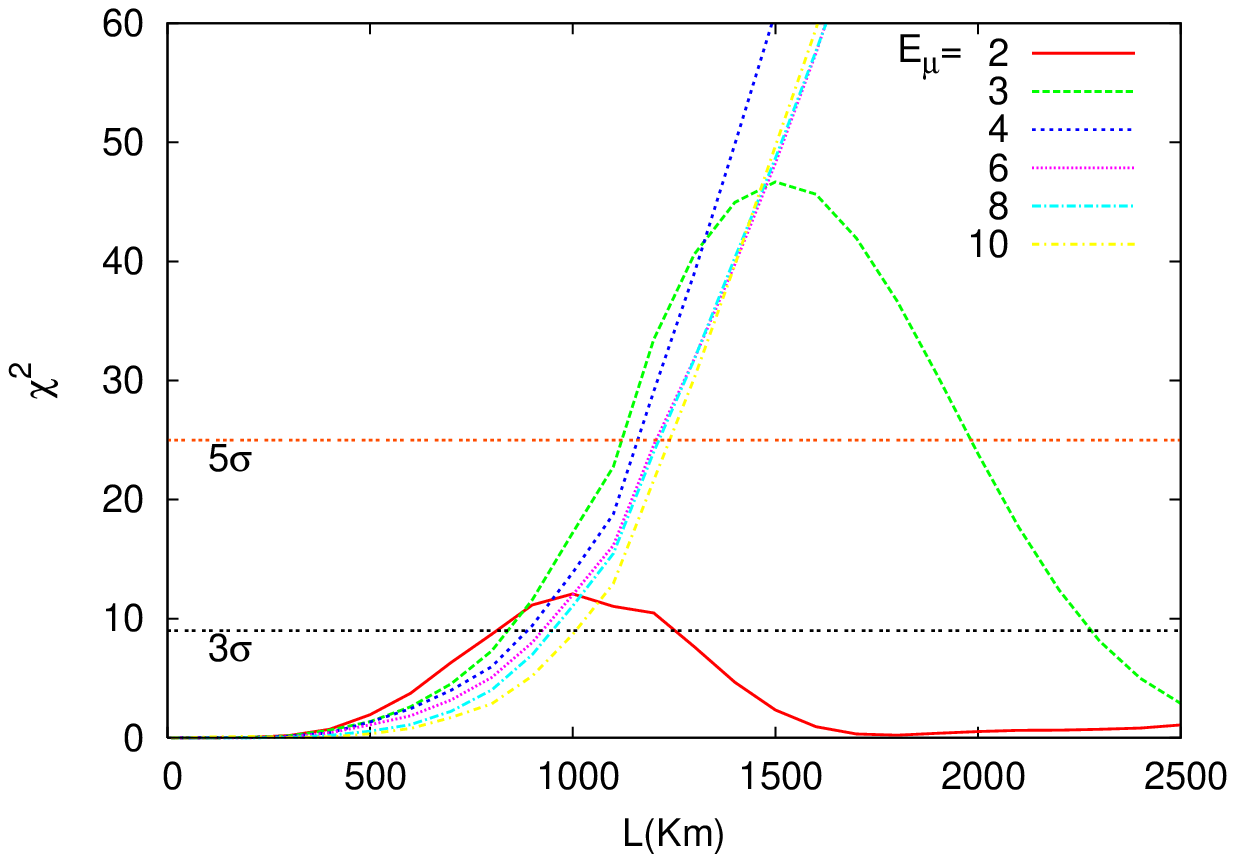}
\end{center}
\caption{\footnotesize{$\chi^2$ as a function of the baseline $L$ for different values of the muon beam energy $E_{\mu}$
at a LENF over a 2 years~(Left) and 5 years~(Right) exposure 
for input $\sin^22\theta_{13}=0.113$ and $\delta_{CP}=90^\circ$. We keep all other
oscillation parameters fixed at their central values and assume normal hierarchy (NH) to be true hierarchy.}}
\label{fig:L-chisq}
\end{figure}

To investigate the hierarchy sensitivity to baselines~($L$), muon beam
energy~($E_{\mu}$), and different detector characteristics such as
energy resolution, efficiency etc, requires a complex numerical
optimization.  In order to find the optimal baseline and the muon beam
energy in determining the mass hierarchy, we assume NH to be the true
hierarchy. We fix our $\sin^2 2\theta_{13}$ in the $3\sigma$ range of
the current best fit value.  We have analysed the $\chi^2$ for
different values of baselines in the range~$250 - 2500\,{\rm km}$ and
for different values of $E_{\mu}$.  We have performed this analysis
for the LENF setup with a running time of $2$ and $5$ years for the
parent muon beam energy $E_{\mu}$ starting at $2.0\,{\rm GeV}$ and
going up to $10.0\,{\rm GeV}$.  We keep all oscillation parameters
fixed at their central values, while $\delta_{CP} = 90^{\circ}$. We
have included the charge mis-identification background at the level of
$10^{-2}$.  In Fig.~\ref{fig:L-chisq}, we use the central value of
$\sin^2 2\theta_{13} = 0.113$ measured by RENO as our true value.  The
two horizontal lines in the plot indicate the $\chi^2$ corresponding
to $3\sigma$ and $5\sigma$. From this figure one
notices that with only 2 years of data, the minimum baselines required
for exclusion of the wrong hierarchy at $5\sigma$, vary from
$1230\,{\rm km}$ to $1600\,{\rm km}$, corresponding to $3-10\,{\rm
  GeV}$ muon beams and that with increased data from 5 years, the
spread in the minimum baselines required reduces appreciably,
particularly for higher energies, resulting in baselines in the range
$1120-1250\,{\rm km}$ being sufficient for a mass hierarchy
discrimination for the same $3-10 \,{\rm GeV}$ range.  Thus as long as
we have $E_{\mu} > 3.0 \,{\rm GeV}$ and $L > \sim 1200\,{\rm km}$,
hierarchy determination is possible at about $5\sigma$ using the
electron disappearance channel alone.

\begin{figure}[htb]
\begin{center}
\includegraphics[width=0.45\textwidth]{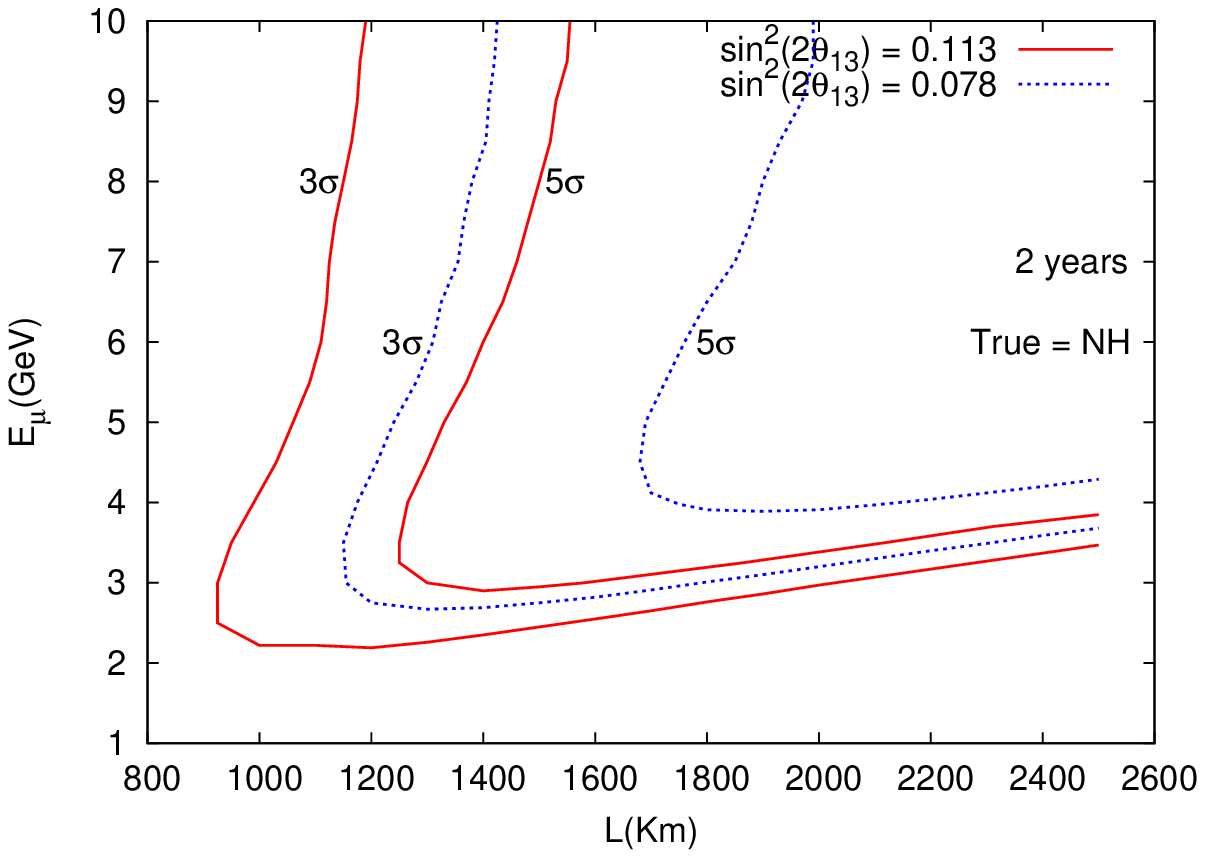}
\includegraphics[width=0.45\textwidth]{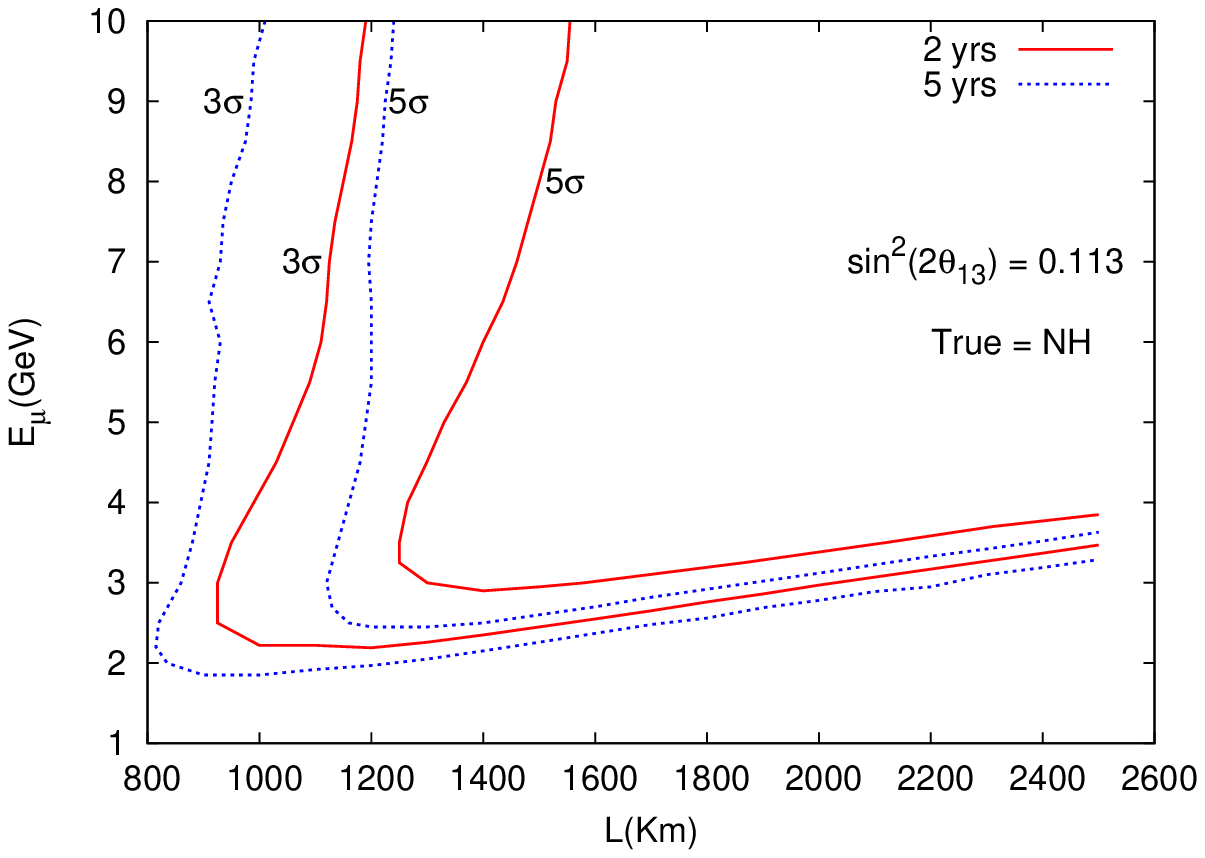}
\end{center}
\caption{\footnotesize{Left:~$3\sigma$ and $5\sigma$ contours in $L - E_{\mu}$ plane at a LENF over a 2 years exposure
for input $\sin^2(2\theta_{13}) = 0.113$~(red/solid curves) and $\sin^2(2\theta_{13}) = 0.078$~(blue/dashed curves), respectively.
Right:~$3\sigma$ and $5\sigma$ contours in $L - E_{\mu}$ plane at a LENF over a 2 years~(red/solid curves) and a 5 years~(blue/dashed curves) exposure
for input $\sin^2(2\theta_{13}) = 0.113$. We have assumed normal hierarchy (NH) to be the true hierarchy.}}
\label{fig:L-emu}
\end{figure}

We next plot the contours correponding to 3$\sigma$ and 5$\sigma$ mass
hierarchy dicrimination in the $E_\mu-L$ plane and these are shown in
Fig.~\ref{fig:L-emu}.  For these plots we marginalize over all the
oscillation parameters. The true value of $\sin^2 2\theta_{13} =
0.113\,\,{\rm and}\,\,0.078$ have been used in the left panel.  While
the first value is the best fit value from RENO, the second smaller
value is 2$\sigma$ lower limit of the global fit of the neutrino
oscillation parameters.  The left figure shows the variation of the
contours with change in the magnitude of $\sin^2 2\theta_{13}$,
whereas the right figure shows the variation for a fixed value with
change in exposure. Again, higher exposure is more effective in
reducing the minimum baseline than the minimum energy, required for a
$3\sigma /5\sigma$ mass hierarchy discrimination. With 5 years data,
beyond $\sim 5$ GeV, there is negligible change in the minimum
baseline needed for a $5\sigma$ mass hierarchy
determination. Moreover, the marginalization of parameters seems to
slightly further reduce the spread in minimum baselines for the $3-10
\,{\rm GeV}$ muon beam energies, at which this measurement is
feasible. While we have shown the contours corresponding only to the
normal hierarchy being the true one, however, similar results also
hold for the case of inverted hierarchy being true, with minor changes
in the values of minimum baseline and muon beam energy at which a
$3\sigma /5\sigma$ mass hierarchy discrimination is achieved.

We would like to mention that since our studies indicate that even
with muon beam energies as low as $3 \,{\rm GeV}$, electron
disappearance can be used for mass hierarchy determination, hence in
principle, this could also be done at the proposed Very low energy
neutrino factory (VLENF)~\cite{BROSS}, in conjunction with an
additional far detector. The VLENF is being proposed as a setup for a
short baseline experiment to resolve the several observed anomalies
which could possibly arise from eV-scale sterile
neutrinos~\cite{Tunnell}. In practise however, for measurements at a
detector $1200 \,{\rm km}$ away, as the neutrino beam will have to be
pointed towards it, this would require that the decay ring for nuSTORM
be built on a slope and the near detector located underground, leading
to huge additional construction costs~\cite{AB}. Moreover, the flux at
VLENF will be few orders of magnitude smaller than that achievable at a
LENF and hence this option does not seem feasible.

\section{Conclusion}
\label{con}
Precision measurement of all the oscillation parameters is the main
goal of future neutrino factories.  Although some parameters are
measured to a very good accuracy, there are still some unknowns such
as the Dirac CP phase $\delta_{CP}$, neutrino mass hierarchy and the
octant of the atmospheric angle $\theta_{23}$. Recent measurements by
Daya Bay, Double Chooz, and RENO have confirmed a large value of
$\theta_{13}$.  This leads to a rather optimistic scenario for
determination of all these unknown parameters.

In this paper, we explore the possibility of using the electron
disappearance channel, i.e, the $\nu_e~(\bar{\nu}_e) \to
\nu_e~(\bar{\nu}_e)$ oscillation, at a low energy neutrino factory
(LENF) to determine the neutrino mass hierarchy.  While the $\nu_e \to
\nu_{\mu}$ golden channel, has been extensively studied for
determination of parameters at LENF, the electron disappearance
channel has the advantage of being independent of the other unknowns:
CP violating phase $\delta_{CP}$ and octant of $\theta_{23}$.  We find
that for muon beam energies $E_{\mu} > 3.0 \,{\rm GeV}$ and baselines
$L > 1200\,{\rm km}$, the electron disappearance channel has the
capability of neutrino mass hierarchy determination at $5\sigma$, for
an exposure of a $20\,{\rm kton}$ totally active scintillating
detector to two years of each muon polarity, with $1.4\times10^{21}$
useful muons decays per year.
 
\acknowledgments 
N.S. would like to thank Alan Bross for a useful discussion regarding VLENF. 
S.K.R. would like to acknowledge WHEPP-XII, where some initial
discussions leading up to this work were carried out.

%%%%%%%%%%%%%%%%%% REFERENCES %%%%%%%%%%%%%%%%%%%%%%%%%%%%

\end{document}